\documentstyle[12pt]{article}
\begin{document}
%----------------------------------------------------------------
\pagestyle{empty}
\parindent 0pt
\null\vskip-40pt
\vfill
\centerline{\bf Critical behaviour of the 1D q-state Potts model}
\vskip .1in
\centerline{\bf with long-range interactions}
\vskip .25in
\centerline{Z Glumac and K Uzelac}
\vskip .25in
{\it Institute of Physics, University of Zagreb, Bijeni\v cka 46, POB 304
, 41000 Zagreb, Croatia}
%\vfill
\vskip 2.5 cm
\centerline{\bf Abstract}

\baselineskip 21pt

The critical behaviour of the one-dimensional q-state Potts model with
long-range
interactions decaying with distance r as $r^{-(1+\sigma)}$ has been
studied in the wide
range of parameters $0 < \sigma \le 1$ and $\frac{1}{16} \le q \le 64$.
A transfer matrix has been constructed for a truncated range of
interactions for integer and
continuous q, and finite range scaling has been applied. Results for
the phase diagram and the
correlation length critical exponent are presented.

\vskip 2cm
Short title: Critical behaviour of the 1D LR q-state Potts model

Physics Abstracts classification number: 0550

\vfill
\eject
\pagestyle{plain}
\parindent 6mm
%   pocetak texta
%-------------------------------------------------------------------------
\section{Introduction}
Recently, finite range scaling (FRS) appeared as an efficient method to deal
with
various discrete one-dimensional models with long range (LR) interactions
(Glumac and Uzelac, 1989, 1991). The aim of the present study is to apply it to
the
one-dimensional
q-state Potts model with LR interactions, defined by the hamiltonian
\begin{equation}
-{\beta}H=K\sum_{i<j}\frac{1}{(j-i)^{1+\sigma}} [\delta(
s_{i},s_{j})-\frac{1}{q}],
\end{equation}
where the interaction constant is chosen to be equal to $k_{B}$ and the
inverse temperature
is denoted by K. The symbol $s_{i}$ denotes a Potts variable at site i
which can assume q
different states, while $\delta(s_{i},s_{j})$ is the Kronecker symbol.

Such a model is expected to have, due to the LR of interactions, a nontrivial
phase
transition at finite temperature, like an Ising model (as its special
case q=2), although
the problem becomes much more complex due to possibility of changing
degrees of freedom
through parameter q.

The Potts model has been extensively studied in the case of short range
(SR) interactions. In
the two dimensional SR case for example, besides integer-q models which
are equivalent to the
classical (q-1)/2-spin models, there are also
non-integer q models which are connected with other critical problems such as
percolation (q=1 limit), resistor network (q=0 limit),
dilute spin glass (q=1/2 limit), branched polymers ($0{\leq}q\leq1$)
 (see review by Wu, 1982).

It would be interesting to extend those concepts to LR interactions. However,
in
contrast to the rich literature on the SR Potts model, the LR Potts
model has been much less
explored. The best explored member of the q-state Potts models is the
q=2 model where both,
analytical (Dyson 1969; Fisher, Ma and Nickel, 1972; Kosterlitz, 1976)
and numerical (Nagle
and Bonner, 1970; Glumac and Uzelac 1989,1991) results are available.
Dyson has shown that,
as a consequence of the LR forces, there is a critical temperature
different from zero as
long as $0 < \sigma \le 1$. Fisher et al. have performed $\epsilon=2\sigma-d$
and 1/n expansions in the context of the renormalization group (RG)
for all dimensions d and $\sigma\ne2$ on a more general system with an
n-component
order parameter. They obtained the correlation function exponent
$\eta=2-\sigma$ to all orders in $\epsilon$,
$\nu=1/\sigma$ when $\epsilon<0$ (classical regime),
$\nu=1/\sigma+\frac{n+2}{n+8}\frac{
\epsilon}{\sigma^{2}}+\dots$ when $\epsilon>0$,
and leading irrelevant exponent $y_{3}=-|\epsilon|$.
A similar expansion, in 1-$\sigma>0$, has been done by Kosterlitz (1976), who
obtained : $1/\nu=-y_{3}=\sqrt{2(1-\sigma)}$ when $\sigma\rightarrow1$.
Very recently, a
new method, called cycle expansion, has been demonstrated by Mainieri (1992)
on the example of the s=1/2 Ising model. Mainieri obtained very
accurate estimates of
the critical temperature and other thermodynamic quantities in the critical
region.

The literature on $q\ne2$ LR Potts models is relatively poor.
By using the RG $\epsilon$-expansion, Priest and Lubensky (1976) have found
a fixed point  and critical exponent $\nu$ to the first order in
$\epsilon=3\sigma-d$. More recently, Theumann and Gusm\~ ao (1985),
performed similar calculation to the second order of $\epsilon=3\sigma-d$.
They obtained the following results for q$<3$: $\eta=2-\sigma$
to all orders of $\epsilon$, and
\begin{equation}
\nu^{-1}=\sigma-\frac{\alpha}{2\beta}\epsilon+\frac{\alpha\beta(\alpha-\beta)
S(\sigma)+\frac{1}{2}\alpha(\beta^{2}-\gamma)G(\sigma)}{8\beta^{3}}\epsilon^{2}
+0(\epsilon^{3})
\end{equation}
where $\alpha=q^{2}(q-2)$, $\beta=q^{2}(q-3)$, $\gamma=q^{4}(q^{2}-6q+10)$,
$S(\sigma)\equiv \psi(\frac{3}{2}\sigma)-\psi(\sigma)-\psi(\sigma/2)+\psi(1)$,
 and $G(\sigma)\equiv \frac{\Gamma(\frac{3}{2}\sigma)\Gamma^{3}(\sigma/2)}
{\Gamma^{3}(\sigma)}$. $\Psi(z)$ is the logarithmic derivative of the $\Gamma$
 function. In the $q\rightarrow1$ limit, their results are applicable
to the description
of percolation critical behaviour for a random Ising ferromagnet with a LR
power-law
interaction.

The advantage of the FRS method is the capability to cover,
unlike the $\epsilon$-expansion,
a wide range of parameter space. In the present article the phase
diagram and the correlation
length critical exponent will be studied in the region $0 < \sigma \le 1$ and
$\frac{1}{16} \le q \le 64$.

The main idea of our approach is the following:
we truncate the originally infinite-range interaction to the L first neighbours
and
solve exactly (although numerically) the finite-range version of (1)
\begin{equation}
-{\beta}H=\sum_{i=1}^{N}\sum_{j=1}^{L}K_{j}
[\delta(s_{i},s_{i+j})-\frac{1}{q}]_{p.b.c.}
\end{equation}
where $K_{j}$ denotes $\frac{K}{j^{1+\sigma}}$.
Then, by using the FRS method (section 2.1) with the appropriate
extrapolating procedure
(section 3), the L$\rightarrow\infty$ behaviour will be deduced.

For this purpose a transfer matrix for a reduced model defined by eq. (3) will
be
constructed. Two different formalisms will be applied: one
for integer q values (section 2.2) and the other for non-integer q values (
section 2.3).

The results are presented in section 3, while section 4 contains a
short discussion and some
open questions.

%-------------------------------------------------------------------------
\section{Method}

\subsection{FRS}

The FRS method has been constructed in analogy with the FSS (Fisher and
Barber, 1972),
where instead of finite-size, the finite-range of interactions
is scaled (Glumac and Uzelac, 1989,1991).
The basic idea of FRS is that by studying the sequence of systems with their
long-range interactions truncated above certain range, one can obtain, by using
scaling properties, the information on the critical behaviour of the true
infinite
system.

Let $A_{\infty}$(t) be the physical quantity of the true long-range system,
which algebraically diverges, in the vicinity of the critical point t=0
\begin{equation}
A_{\infty}(t)\simeq A_{0}t^{-\rho}
\end{equation}
where t=(T-$T_{c}$)/$T_{c}$, $T_{c}$ is the critical temperature,
$\rho$ is the related critical exponent, and $A_{0}$ is a constant.
Then, analogous to the FSS hypothesis one can assume
 that for large finite range M and small t, $A_{M}$(t)
can be written in the form
\begin{equation}
A_{M}(t)=A_{\infty}(t){\cdot}f(\frac{M}{\xi_{\infty}})
\end{equation}
where f is a homogeneous function with following properties
\begin{equation}
\lim_{x\rightarrow\infty}f(x)=  1, {\hskip 2cm}
\lim_{x\rightarrow0}f(x)=const.{\cdot}x^{\frac{\rho}{\nu} }
\end{equation}

By applying the equation (5) to the correlation length $\xi_{\infty}(t)=
\xi_{0}t^{-\nu}$, the standard scaling procedure gives the condition for
critical
temperature through the fixed point equation
\begin{equation}
\frac{\xi_{M}(t^{\ast})}{M}=\frac{\xi_{M'}(t^{\ast})}{M'}
\end{equation}
and the expression for correlation length critical exponent $\nu$
\begin{equation}
\nu^{-1}=\frac{ \ln\frac{\xi'_{M}(t^{\ast})}{\xi'_{M'}(t^{\ast})}}
{\ln\frac{M}{M'}}-1
\end{equation}
where M'=M-1 in all calculations,
and $\xi$' is temperature derivative of $\xi$.

There are two important facts concerning the above method which are of
interest.

First, since the
critical behaviour is essentially dependent on the range of interaction, one
can (unlike in the FSS (Br\' ezin, 1982))
expect applicability of FRS in mean-field as well as in the non-trivial
 region (Glumac and Uzelac 1989, 1991).
Second, the correlation length calculated from the transfer matrix
presents the average
distance between domain walls for an infinitely long strip so that
relation (7) gives the
transition temperature for both first and second-order phase transitions.
This fact is common with FSS
(Binder, 1987).
%-------------------------------------------------------------------------

\subsection{Transfer matrix: integer-q formalism}

It is straightforward to construct the transfer matrix for model (3)
with an integer number of Potts states.

The chain with a range of interaction L, can be represented as a strip
with columns of
height L (Uzelac and Glumac, 1988).
Each column then can be considered as an object with q$^{L}$ possible states,
interacting only with its first neighbour.

The transfer matrix {\bf T} is given by:
\begin{equation}
\langle i \vert {\bf T} \vert j \rangle = \exp \left \{ \sum_{k=1}^{L}
K_{k}\left[ \sum_{n=1}^{L-k} \delta(i_{n},i_{n+k}) +
\sum_{n=1}^{k} \delta(i_{L+n-k},j_{n})\right] \right \}
\end{equation}
where $j_{m}=0,1,\dots,q-1$ are the elements of the L-component vector of
states $\vert j \rangle$ of a column of height L.
The matrix {\bf T} can be further decomposed
into a product of L matrices {\bf T}$_{n}$,
each one describing the addition of one more site to the column
(Temperley and Lieb, 1971).
\begin{equation}
{\bf T}={\bf T}_{1}\cdot\dots\cdot {\bf T}_{L}
\end{equation}
\begin{equation}
\langle i \vert {\bf T}_{n} \vert j \rangle = \prod_{l=1 \atop l\ne n}^{L}
\delta(i_{l},j_{l}) \exp \left \{ \sum_{k=1}^{n-1} K_{k}\delta(j_{n-k},j_{n}) +
\sum_{k=n}^{L-1} K_{k} \delta(j_{L+n-k},j_{n}) + K_{L}\delta(i_{n},j_{n})
\right \}
\end{equation}
 There is a simple relation between neighbouring one-site matrices:
\begin{equation}
{\bf U}^{T}{\bf T}_{n+1}{\bf U}={\bf T}_{n}
\end{equation}
where {\bf T}$_{L+1}$={\bf T}$_{1}$ and {\bf U} is a matrix of the
translation operator in the vertical strip direction, with matrix elements
given by
\begin{equation}
\langle i \vert {\bf U}\vert j \rangle=\delta(i_{1},j_{L})
\delta(i_{2},j_{1}) \delta(i_{3},j_{2}), \dots, \delta(i_{L-1},j_{L-2})
\delta(i_{L},j_{L-1}) .
\end{equation}
The matrix {\bf U } satisfies the relations
${\bf U}^{L}={\bf 1}$ and ${\bf U}^{L-1}={\bf U}^{T}={\bf U}^{-1}$.
Consequently, the transfer matrix can be written as the L-th power of a
single matrix
as a peculiarity of the present long-range model:
\begin{equation}
{\bf T}=({\bf U}\cdot{\bf T}_{L})^{L}=\tilde{\bf T}^{L}
\end{equation}
\begin{equation}
\langle i \vert \tilde{\bf T}\vert j \rangle=\delta(j_{1},i_{2})
\delta(j_{2},i_{3})
\dots\delta(j_{L-1},i_{L})\exp \left \{ \sum_{m=1}^{L}K_{L+1-m}
\delta(j_{L},i_{m}) \right \}
\end{equation}
Notice that \~{\bf T} has only q nonzero elements per row which
greatly reduces computer memory.

Further reduction
is obtained by using the symmetry properties of
Potts interaction. In the present calculation, only the translation
invariance in the
space of Potts states has been used, which decomposes \~{\bf T}
into q submatrices (of order $q^{L-1}$). The q-1 of them have
the identical eigenvalues as a consequence of Kronecker-$\delta$ type
of interaction.
Thus we have to diagonalize only two submatrices or, more precisely, we have to
find
 only the largest eigenvalue of each of the two submatrices,
$\mu_{1,L}$ and $\mu_{2,L}$
respectively, to be able to construct the correlation length $\xi_{L}$
\begin{equation}
\xi_{L}=\frac{L}{\ln\frac{\lambda_{1,L}}{\lambda_{2,L}}}=
\frac{1}{\ln\frac{\mu_{1,L}}{\mu_{2,L}}}
\end{equation}
where $\lambda1>\lambda2$ and $\mu1>\mu2$ are the eigenvalues of
{\bf T} and \~{\bf T} respectively.

For this purpose, the method of direct iteration with a single vector is
suitable (Wilkinson, 1965). If one defines a set of vectors
{\bf b}$_{j}$ in a following way
\begin{equation}
{\bf A}{\bf b}_{j}={\bf b}_{j+1}
\end{equation}
for arbitrary initial value {\bf b}$_{0}$, where {\bf A} is a real nonsymmetric
matrix, then the largest eigenvalue $\mu$ of matrix {\bf A} is given by
\begin{equation}
\mu=\lim_{j\rightarrow\infty}
\frac{b_{j}(n)}{b_{j-1}(n)}
\end{equation}
for any vector's component n. By applying this method for cases with q= 2,
3, 4 and 5, we obtained the two largest eigenvalues
 of \~{\bf T} for maximum ranges L= 20, 13, 10 and 9, respectively.
A few ten to a few hundred iterations were required to obtain the leading
eigenvalue with an accuracy of $10^{-15}$ for both submatrices.

%-------------------------------------------------------------------------
\subsection{Transfer matrix: continuous-q formalism}

As mentioned in Introduction, there are some interesting cases (q=0, 1/2, 1,
...)
to which the described integer-q formalism cannot be applied or becomes
inefficient
(limitation to small ranges for large q models).
Thus, it would be advantageous to formulate a q-independent
transfer matrix, which would allow to treat q as a continuous parameter.
Such a type of
transfer matrix has already been constructed for the 2D Potts model with
SR interaction (Bl\"ote, Nightingale, Derrida 1981;
Bl\"ote and Nightingale, 1982).
We will shortly explain the application of that method to the considered LR
model.

%   napisati partic funkc. preko grafova

Let us translate the partition function of model (3), into the
graph-theory language.
First, we rewrite the basic expression $\exp\{K_{j}\cdot\delta
(s_{i},s_{i+j})\}$ as
\begin{equation}
e^{K_{j}\delta(s_{i},s_{i+j})}=1+v_{j}\delta(s_{i},s_{i+j}),
\end{equation}
where $v_{j}=e^{K_{j}}-1$, so that the partition function, with periodic
boundary
conditions, becomes
\begin{equation}
Z_{N,L}=\sum_{s_{1}=1}^{q}\sum_{s_{2}=1}^{q}\dots\sum_{s_{N}=1}^{q}
\prod_{j=1}^{L}
\end{equation}
\begin{displaymath}
[1+v_{j}\delta(s_{1},s_{1+j})][1+v_{j}\delta(s_{2},s_{2+j})]\dots
[1+v_{j}\delta(s_{N},s_{N+j})]
\end{displaymath}

If we perform the multiplications of the above terms, the one-to-one
correspondence between every term of that expansion and the particular graph on
the m$\times$L lattice (N=m$\cdot$L) can be established in the standard way
(Bl\"ote and Nightingale, 1982).

Now, the partition function becomes
\begin{equation}
Z_{N,L}=q^{N}\cdot\sum_{G}u^{b_{1}(G)}_{1}{\cdot}u^{b_{2}(G)}_{2}\cdot\dots
{\cdot}u^{b_{L}(G)}_{L}{\cdot}q^{l(G)}.
\end{equation}
The above summation goes over all possible graphs on the lattice with L
different types of
bonds described by $u_{j}=v_{j}/q$
, $b_{j}$(G) is the number
 of bonds of type j, and l(G) is the number of independent loops on the graph
G.

%It is very important to notice that q appears only as a parameter in the above
%expression, so he is by no means restricted to the integer values.

%              povezanost i TM

A certain number of different bonds in every single graph, produce
connections between sites. If one considers a strip
of height L, then all possible interactions between all sites in that strip
produce
a particular state of connectivity $\alpha$ of the sites in the last column.
The LR interactions produce a larger number of connectivities than in
the 2D SR models.

The connectivity is represented by
an integer, and it will be used as an index of the transfer matrix.
The partition function Z$_{N}$ can be written as a vector with components
Z$_{N}$($\alpha$)
, where each component Z$_{N}$($\alpha$) describes the contribution
to the partition function of all graphs that
produce one particular state of connectivity $\alpha$ in the last column.
By adding a new column, we get Z$_{N+L}$ which again can be written as a vector
 Z$_{N+L}$($\beta$). The connection between those two partition
functions is given by
the transfer matrix in the following way
\begin{equation}
Z_{N+L}(\beta)=\sum_{\alpha} {\bf T}(\beta,\alpha)Z_{N}(\alpha)
\end{equation}
where the summation goes over all graphs, produced by the added column,
that lead from
connectivity $\alpha$ to $\beta$.
A transfer matrix decomposition analogous to that of section 2.1, leads to
\begin{equation}
{\bf T}=q{\bf T}_{1}\cdot\dots\cdot q{\bf T}_{L}=(q{\bf T}_{1}{\bf V}
^{T})^{L}=\tilde{\bf T}^{L}
\end{equation}
where ${\bf T}_{j}$ describes the addition of the j-th site to the column, and
{\bf V} is the matrix of the translation operator in the vertical strip
direction.
Further decomposition of {\bf T}$_{1}$ is possible
\begin{equation}
{\bf T}_{1}={\bf T}_{1,L}(u_{L})\cdot\prod_{j=L-1}^{1}{\bf T}_{1,j}(u_{j})
\end{equation}
Each of the {\bf T}$_{1,j}$(u$_{j}$) matrices describes
the presence of an u$_{j}$ type bond
 on the first site and has at most two nonzero elements per row.
{\bf T}$_{1,L}$(u$_{L}$) is an upper triangular matrix, while the other
matrices are
lower triangular. It is easy to construct a matrix {\bf M} which relates the
lower
 triangular matrices
\begin{equation}
{\bf M}^{T}{\bf T}_{1,j}(x){\bf M}={\bf T}_{1,j+1}(x)
\end{equation}
The matrix \~{\bf T} then becomes
\begin{equation}
\tilde{\bf T}=q{\bf T}_{1,L}(u_{L})\prod_{j=L-1}^{1}({\bf T}_{1,L-1}
{\bf M})_{u_{j}}{\bf V}^{T}
\end{equation}

By using these decompositions, it is sufficient to keep in
computer memory only four matrices:
{\bf T}$_{1,L}$, {\bf T}$_{1,L-1}$, {\bf M} and {\bf V}, with at most two
nonzero elements per row, which strongly reduces amount of data to be stored.

% izbaciti ovaj dio ????
% Let us only mention that the matrices {\bf M} and {\bf V} are connected by
%%the
% real-space reflection operator {\bf R} in the following way
% \begin{equation}
% {\bf M}={\bf R} \cdot {\bf V} \cdot {\bf R}
% \end{equation}
% where matrix of {\bf R} is given by
% \begin{equation}
% \langle i \vert {\bf R} \vert j \rangle = \delta(i_{L},j_{1})
% \delta(i_{L-1},j_{2}) \dots \delta(i_{1},j_{L})
% \end{equation}

Since we want to get information about LR correlations, we should also
take care of the connectivity between sites in the last column and the
sites in the first
column. In order to
do so, one can imagine the top site in the last column of Z$_{N}$
to be some kind of "ghost" site representing the connection between the
first site of
last column of Z$_{N+L}$ and the sites of the first column.
Thus we need a column of height L to describe all connectivities in the model
with
 the range of interaction equal to L-1. In order to keep the "ghost"
site fixed, the
matrix of the translation operator {\bf V} will be constructed by acting with
operator {\bf V} on non-"ghost" sites only.
Consequently, if we perform the following set of transformations
\begin{equation}
u_{L}\rightarrow u_{L-1}, u_{L-1}\rightarrow u_{L-2}, \dots,
u_{2}\rightarrow u_{1}, u_{1}\rightarrow 0,
\end{equation}
we will obtain \~{\bf T} for the model with range L-1, with LR correlations
included. With this redefinition, \~{\bf T} has the following structure
%     ispis matrice
%\begin{equation}
\[\tilde{\bf T}=\left(\matrix{\tilde{\bf T}_{-}&{\bf X}\cr
{\bf 0}&\tilde{\bf T}_{+}\cr}\right)\]
%\end{equation}
where \~{\bf T}$_{+}$ and \~{\bf T}$_{-}$ contain the largest $\mu_{1,L}$
and the second largest $\mu_{2,L}$ eigenvalue of the whole matrix \~{\bf T},
respectively. The correlation length is given by the relation (16) again.

The order of \~{\bf T} for a model with range L
is given by N$_{L}$, and the order of \~{\bf T}$_{+}$ is given by N$_{L-1}$,
where
\begin{equation}
N_{L}=\sum_{i=1}^{L+1}\varphi_{i}^{L}
\end{equation}
\begin{equation}
\varphi_{i}^{L}=i\cdot\varphi_{i}^{L-1}+\varphi_{i-1}^{L-1}
\end{equation}
for L$\geq$2, and $\varphi_{2}^{2}$=3, $\varphi_{1}^{n}=\varphi_{n+1}^{n}=1$.
Some values of N$_{L}$ are presented in table 1.

The present calculations were performed for the maximum range L=8 with
q=1/16, 1/2, 1, 8, 16, 32 and 64. The largest eigenvalues of the
matrices \~{\bf T}$_{\pm}$, are extracted by using the method
 explained earlier (section 2.2) with a similar accuracy and number of
iterations.

%-------------------------------------------------------------------------
\section{Results}
\subsection{Convergence and extrapolation procedures}

The critical temperature and exponent $\nu$, given by the equations (7) and
(8), still depend on the range M and it is important to apply the right
extrapolation procedure.
A simple analysis done within FSS (Privman and Fisher, 1983) can be
reproduced in the case
of the FRS (Glumac and Uzelac, 1989) and shows that the convergence is
dominantly governed by
the leading irrelevant field and the corresponding critical exponent
$y_{3}<0$. On the
basis of such analysis, we expect power-law
convergence of T$_{c}$ and $\nu$ in the large-M limit
\begin{equation}
T_{c,M}=T_{c}+a\cdot M^{y_{3}-\frac{1}{\nu}}
\end{equation}
\begin{equation}
\nu_{M}^{-1}=\nu^{-1}+ b(T_{c,M}-T_{c})\cdot M^{\frac{1}{\nu}}
+c\cdot M^{y_{3}}
=\nu^{-1}+b' \cdot M^{y_{3}}
\end{equation}
where a, b, $b'$ and c are constants. Thus, we chose to fit the
obtained results for
$K_{c,M}$ and $\nu_{M}$ to the form
\begin{equation}
y_{M}=y_{e}+A\cdot M^{-x_{y}}
\end{equation}
in the least-squares approximation (LSA), where $y_{e}$ denotes the
extrapolated
quantity. The calculations are performed by taking the five largest values of
M.

It should be noticed, however, that Privman and Fisher (1983) analysis
does not apply to
the MF
region and, consequently, the extrapolation form (32) becomes arbitrary
there. On the other
hand, in our previous analysis for the Ising model (Glumac and Uzelac,
1989) it turns out that
the extrapolation form
\begin{equation}
y_{M}=B+A(\frac{M-1}{M})^{x_{y}}
\end{equation}
gives a better agreement with the exactly known results for $\nu$ in the MF
region
then the function of the form (32).

The error bars of the above extrapolations can be estimated only
roughly, by looking for the
remaining L dependence of the results.
In tables (2) and (3) are presented results which last digit is
modified under the change of
maximal range from L-1 to L. The more careful examination of the
errors, shows that the size of
the error estimate changes with q and $\sigma$,
being less then ten percents at the most of q and $\sigma$
region. Exception is the small q and $\sigma$ region where the error is
estimated to be
a few times larger.

%-------------------------------------------------------------------------
\subsection{Critical temperature}

The inverse critical temperature $K_{c,M}$, defined by equation (7), is
calculated
with an accuracy higher than $10^{-10}$
for different values of Potts states and with $\sigma$ as an parameter.

When increasing the exponent of LR interaction, there will be a value of
$\sigma$
beyond which the SR critical behaviour (with $T_{c}=0$) takes place.
For the Ising model (q=2) it was analytically shown (Dyson, 1969) that this
$\sigma$
 is equal to 1. As argued more generally (Fisher et al. 1972, Sak 1973)
this exchange of regimes should occur when the correlation function exponent
$\eta$ of the LR system becomes smaller or equal to the SR one, $\eta_{SR}$.
Thus in present case, we expect that $\sigma_{c}=1$ for all values of q. This
expectation is confirmed by our numerical results.
In our earlier works (Glumac and Uzelac, 1989, 1991),
we have been able to detect the appearance
of SR forces governed critical behaviour by the change of $K_{c,M}$
from descending to ascending sequence and by the sudden increase of
convergence exponent.
In the present calculations, both of these phenomena were observed for
any q considered
at $\sigma\approx1$. The extrapolated values for the critical
temperature are presented in
figures (1a) and (1b).

We have used the LSA extrapolation method with the form (32) to
calculate $K_{c,e}$ and
the leading convergence exponent $x_{K}$ as a function of q and $\sigma$
which are presented in tables (2a) and (2b) respectively.
The only point where this could not be applied is $\sigma=1$, where the
simple convergence
expression does not apply any more. In order to avoid non-monotonic
behaviour, we have used
only the data for the three largest values of M fitting a linear
relation to them ($x_{K}=1$).
The error should not be important, since for  $\sigma=1$ the variation
of data with M is
rather weak.

%-------------------------------------------------------------------------
\subsection{Critical exponent $\nu$}

The correlation length critical exponents  $\nu_{M}$ has been calculated by
using the equation
(8). The accuracy of results is reduced (with respect to that of the
critical temperature) to the
order of $10^{-6}$ due to the numerical differentiation.

Following the form (31) for the scaling correction in non-MF region,
one can expect that the correction
terms to $\nu_{M}^{-1}$ would be smaller while calculating the
$\nu_{M}$ on the temperature
$K_{c,e}$ instead of $K_{c,M}$.
This is indeed the case for the q=2 (s=1/2 Ising) model, where detailed
comparison between
$\nu_{M}(K_{c,M})$ and $\nu_{M}(K_{c,e})$ has been given (Glumac and
Uzelac, 1989).
In the present model we observe better convergence of the data
calculated from $K_{c,e}$ in the LR $\sigma$-region, for all q.

According to the discussion at the end of section 3.1, the extrapolations in
the
low $\sigma$ region where MF behaviour is expected
were performed at the temperature $K_{c,M}$ using the extrapolation function of
the
form (33). Since the $\sigma_{MF}$ border is still an open question for
$q > 2$ (see later in
text), we have used eq. (33) for $\sigma < .3, q \le 1$, and for
$\sigma < .5, q = 2$.

The extrapolated results are presented in figure (2). In tables
(3a) and (3b) are results for $\nu_{e}^{-1}$ and the corresponding
convergence exponents
respectively.
Similarly to preceding section, for $\sigma=1$ the LSA procedure was performed
by
imposing $x_{\nu}=1$ in eq. (32).

The special case q=2 which corresponds to the Ising model was extensively
studied in our earlier work (Glumac and Uzelac, 1989) with a maximum range
equal 10.
The present work permits to reach the range of 20.
Similar values of $\nu_{e}$ were obtained in
both works, which suggests that an increase of range from 10 to 20 does
not change the accuracy significantly.

On table 4, our results are compared to the $\epsilon$-expansion results of
Theumann and Gusm\~ ao (1985). A difference $\Delta$ of only few
percent is obtained.
% Note, that according to relation (2), all
% exponents for $q<3$ and $\sigma=\frac{1}{3}$ are equal to $\frac{1}{3}$
% (fig. 2a).
Figure (2a) suggests that $\sigma=\frac{1}{3}$ as the MF
border and a q-independent value of $\nu_{MF}^{-1}=\sigma$ for all $q \le 1$.

We have not found any
analytical or numerical estimate for the values of $\sigma_{MF}$ or
$\nu$ for $q \ge 3$.
In that situation we decided to calculate $\nu$ at the temperature
$K_{c,e}$ since the
scaling law $\frac{\xi_{M}}{M}=const.$ is more accurate at that temperature
than at
$K_{c,M}$.
Also, we decided to make the extrapolation using function (32) which posesses a
more
transparent L-dependence.

For large q ($q\ge 16$), only the continuous formalism could be
applied, which does not
permit to go beyond the range of L=8. Contrary to the small q case, for
large q this range
was not sufficient to find a good extrapolation. Namely, when applying
the LSA there, with
the assumed form (31), one obtains $x_{\nu}\rightarrow 0$ so that the
correction terms
become of the same order as the leading term and $\nu^{-1}$ diverges.
This behaviour of $\nu$ suggests a change of regime of convergence and
brings up the question
of the order of transition.

For all values of q the exponent $\nu$ increases as
$\sigma\rightarrow1$, which in the case
q=2 appeared (Glumac and Uzelac, 1989) as an indication of the
essential singularity
in this limit (Kosterlitz 1976).

%-------------------------------------------------------------------------
\section{Conclusion and discussion}

By using a FRS method combined with transfer matrix calculations we
have been able to study
the long-range Potts model on the one-dimensional lattice in a wide
range of values for
the number of states q and the interaction exponent $\sigma$. We have
shown that for this
problem the transfer matrix can be decomposed into matrices of much simpler
form. A
different transfer matrix procedure can be done also in the case of
continuous q, so that
our analysis could be extended to some non-integer q cases of interest.

The study is concentrated on the phase diagram and the correlation
length critical exponent
$\nu$.

The critical temperature shows monotonic decrease with q and $\sigma$
and has a finite value
at $\sigma=1$. The change of behaviour of $K_{c,M}$ at $\sigma\approx1$
indicates the
exchange from LR to SR regime, established for any q.
The critical exponent was calculated at the extrapolated critical temperature,
and good agreement with values obtained by other authors was observed.
Generally,
a good accuracy is harder to obtain in the small-$\sigma$ region due to
the very long-range
of interaction and in the $\sigma\rightarrow1$ region where the range of
interaction ceases to be a good scaling variable.

There are two points that require some further discussion.

First is a question of mean-field border. The present method, unlike
the FSS, has the
advantage that it can be applied within the mean-field region, but on
the other hand,
passes smoothly between the two regions (Glumac and Uzelac 1989, 1991)
, which does not
permit to point out the mean-field border $\sigma_{MF}$. Another way to
detect this border
could be numerical, by observing the change of behaviour of $\nu$ as a function
of
$\sigma$. But, our previous results on the Ising model where
$\sigma_{MF}$ is known, show
that close to it the numerical results for  $\nu$ are not sharp enough
to locate this point
with precision. In the present case, we can only confirm that our
results are consistent with
 $\sigma_{MF}$ as known from the literature for a few particular cases:
q=2 ($\sigma_{MF}=
\frac{1}{2}$ Fisher et al 1972), q=1 ($\sigma_{MF}=\frac{1}{3}$ Priest
and Lubensky 1976),
$q=\frac{1}{2}$ ($\sigma_{MF}=\frac{1}{3}$ Aharony 1978, Aharony and
Pfeuty 1979), but
the precise location of $\sigma_{MF}$ for arbitrary q is left unestimated.

A second point is the possible appearance of a first order transition
for some $q>q_{c}$. Within
the transfer matrix formalism this problem was considered for the 2D SR
case where $q_{c}$
exists and is exactly known ($q_{c}=4$, Baxter 1973). Igl\' oi and S\'
olyom (1983) have
shown that the first order transition is connected with the crossing of
the largest and the
third largest eigenvalue of the transfer matrix for finite length chain
and large q.
We have applied a similar calculation in our model considering two groups of
parameters (q=100, $\sigma=.8$, L=4 and q=300, $\sigma=.4$, L=4). Both cases
give a negative result; the first and third eigenvalues do not intersect
as a function of temperature. By analogy with Igl\' oi and S\' olyom, that
result can be interpreted as the absence of a first-order transitions for any q
in the 1D LR Potts model.
Since this is opposite to the indications of the behaviour of $\nu$ for
large q (at the end of section 3.3), the question of the order of
transition for large q is
still left open.
%

%-------------------------------------------------------------------------
\newpage
{\bf References}
% \begin{enumerate}
% \item

Aharony A 1978 J. Phys. C {\bf 11} L457
% \item

Aharony A and Pfeuty P 1979 J. Phys. C {\bf 12} L125
% \item

Baxter R J 1973 J. Phys. C: Solid State Phys. {\bf 6} L445
% \item

Binder K 1987 Rep. Prog. Phys. {\bf 50} 783
% \item

Bl\" ote H W J, Nightingale M P, Derrida B 1981 J. Phys. A: Math. Gen. {\bf 14}
L45
% \item

Bl\" ote H W J and Nightingale M P 1982 Physica {\bf 112A} 405
% \item

Br\' ezin E 1982 J. Physique {\bf 43} 15
% \item

Dyson F J 1969 Commun. Math. Phys. {\bf 12} 91
% \item

Fisher M E and Barber M N 1972 Phys. Rev. Lett. {\bf 28} 1516
% \item

Fisher M E, Ma S K, Nickel B G 1972 Phys. Rev. Lett. {\bf 29} 917
% \item

Glumac Z and Uzelac K 1989 J. Phys. A: Math. Gen. {\bf 22} 4439
% \item

Glumac Z and Uzelac K 1991 J. Phys. A: Math. Gen. {\bf 24} 501
% \item

Hamer C J and Barber M N 1981 J. Phys. A: Math. Gen. {\bf 14} 2009
% \item

Igl\' oi F and S\' olyom J 1983 J. Phys. C: Solid State Phys. {\bf 16} 2833
% \item

Kosterlitz J M 1976 Phys. Rev. Lett. {\bf 37} 1577
% \item

Mainieri R 1992 Phys. Rev. A {\bf 45} 3580
% \item

Nagle J F and Bonner J C 1970 J. Phys. C: Solid State Phys. {\bf 3} 352
% \item

Priest R G and Lubensky T C 1976 Phys. Rev. B {\bf 13} 4159
% \item

Privman V and Fisher M E 1983 J. Phys. A: Math. Gen. {\bf 16} L295
% \item

Sak J 1973 Phys. Rev. B {\bf 8} 281
% \item

Temperley H N V and Lieb E H 1971 Proc. Roy. Soc. Lond. A. {\bf 322} 251
% \item

Theumann W K and Gusm\~ ao M A 1985 Phys. Rev. B {\bf 31} 379
% \item

Uzelac K and Glumac Z 1988 J. Phys. A: Math. Gen. {\bf 21} L421
% \item

Vanden Broeck J M and Schwartz L W 1979 SIAM J. Math. Anal. {\bf 10} 658
% \item

Wilkinson J H The Algebraic Eigenvalue Problem Clarendon Oxford 1965
% \item

Wu F Y 1982 Rev. Mod. Phys. {\bf 54} 235
% \item

% \end{enumerate}

%-------------------------------------------------------------------------
\newpage

Figure captions:
\vskip .25in

Figure 1a:

The extrapolated critical temperature as a function of $\sigma$.
% for:
% $q=\frac{1}{16}$ circle; $q=\frac{1}{2}$ square; q=1 triangle.

\vskip .25in

Figure 1b:

The extrapolated critical temperature as a function of $\sigma$.
Figures (a) and (b) have common q=1 line which allows the comparison of scales
on
both figures.

% for:
% q=1 upright and inverted triangle; q=2 square with plus sign; q=3
solid filled circle;
% q=4 solid filled triangle;
% q=5 solid filled square; q=8 triangle;
% q=16 inverted triangle; q=32 square;
% q=64 circle.

\vskip .25in

Figure 2a:

The exponent $\nu^{-1}_{e}$ as a function of  $\sigma$.
% for:
% $q=\frac{1}{16}$ circle;  $q=\frac{1}{2}$ square; q=1 triangle; q=2
solid filled circle.

\vskip .25in

Figure 2b:

The exponent $\nu^{-1}_{e}$ as a function of  $\sigma$.
Figures (a) and (b) have common q=2 line which allows the comparison of scales
on
both figures.
% for:
% q=2 circle; q=3 square; q=4 triangle; q=5 solid filled triangle;
% q=8  solid filled circle.

%-------------------------------------------------------------------------
\newpage

Table captions:

\vskip .25in

Table 1:

$N_{L}$ is the order of
transfer matrix for continuous-q model with maximum range equal L.

\vskip .25in

Table 2a:

The extrapolated values of inverse critical temperature as a function of q and
$\sigma$. For the error bars see in text.

\vskip .25in

Table 2b:

The critical temperature convergence exponent $x_{K}$ as a function of q and
$\sigma$.

\vskip .25in

Table 3a:

The extrapolated values of $\nu^{-1}$ as a function of q and $\sigma$.
For the error bars see in text.

\vskip .25in

Table 3b:

The convergence exponent $x_{\nu}$ as a function of q and $\sigma$.

\vskip .25in

Table 4:

The comparison between extrapolated values $\nu^{-1}_{e}$ and
Theumann and Gusm\~ ao values $\nu^{-1}_{TG}$ for $\sigma=.4$.

%-------------------------------------------------------------------------

\newpage

\begin{tabular}{||c||c|c|c|c|c|c|c|c|c||} \hline \hline
L & 2 & 3 & 4 & 5 & 6 & 7 & 8 & 9 \\ \hline
$N_{L}$ & 5 & 15 & 52 & 203 & 877 & 4140 & 21147 & 115975 \\ \hline \hline
\end {tabular}

%               -----------  table 1 -------------

\vskip .5in

\begin{tabular}{||c||c|c|c|c|c|c|c|c|c|c|c||} \hline \hline
$\sigma\backslash$ q  & $\frac{1}{16}$ & $\frac{1}{2}$ & 1 & 2 & 3 & 4 & 5 & 8
&
16 & 32 & 64 \\ \hline \hline
.1 & .004 & .031 & .061 & .0927 & .136 & .203 & .28 & .48
& .78 & 1.00 & 1.14 \\ \hline
.2 & .007 & .052 & .102 & .1831 & .2701 & .362 & .45 & .64 & .91 & 1.12 & 1.31
\\ \hline
.3 & .010 & .077 & .149 & .2717 & .3862 & .489 & .576 & .76 & 1.02 & 1.24 &
1.45 \\
 \hline
.4 & .014 & .108 & .204 & .3625 & .4939 & .601 & .690 & .875 & 1.14 & 1.37 &
1.59 \\ \hline
.5 & .020 & .147 & .270 & .4590 & .6013 & .713 & .803 & .99 & 1.26 & 1.53 &
1.75
\\ \hline
.6 & .028 & .198 & .351 & .5644 & .7143 & .829 & .920 & 1.12 & 1.40 & 1.69
& 1.94 \\ \hline
.7 & .041 & .269 & .452 & .6833 & .8374 & .954 & 1.046 & 1.24 & 1.54 & 1.84
& 2.15 \\ \hline
.8 & .064 & .375 & .584 & .8231 & .9774 & 1.093 & 1.185 & 1.384 & 1.69 & 2.00
 & 2.336 \\ \hline
.9 & .121 & .552 & .763 & .9973 & 1.144 & 1.255 & 1.343 & 1.540 & 1.84 &
2.173 & 2.519 \\ \hline
1.0 & .430 & .815 & .990 & 1.230 & 1.348 & 1.440 & 1.518 & 1.697 & 1.997 &
2.325 & 2.677 \\ \hline \hline
\end{tabular}
%                ------------ table 2a ------------

\vskip .5in

\begin{tabular}{||c||c|c|c|c|c|c|c|c|c|c|c||} \hline \hline
$\sigma\backslash$ q  & $\frac{1}{16}$ & $\frac{1}{2}$ & 1 & 2 & 3 & 4 & 5 & 8
&
16 & 32 & 64 \\ \hline \hline
.1 & 1.5 & 1.4 & 1.3 & .86 & .81 & .8 & .9 & 1.0 & 1.1 & 1.0 & .8 \\ \hline
.2 & 1.5 & 1.4 & 1.3 & 1.02 & .99 & 1.1 & 1.1 & 1.2 & 1.2 & 1.0 & .9 \\ \hline
.3 & 1.6 & 1.5 & 1.4 & 1.13 & 1.13 & 1.2 & 1.3 & 1.3 & 1.2 & 1.0 & .9 \\ \hline
.4 & 1.6 & 1.5 & 1.4 & 1.19 & 1.21 & 1.3 & 1.3 & 1.3 & 1.2 & 1.0 & .9 \\ \hline
.5 & 1.6 & 1.4 & 1.3 & 1.22 & 1.25 & 1.3 & 1.4 & 1.3 & 1.2 & 1.1 & .9 \\ \hline
.6 & 1.6 & 1.4 & 1.3 & 1.22 & 1.26 & 1.3 & 1.4 & 1.4 & 1.3 & 1.2 & 1.0 \\
\hline
.7 & 1.5 & 1.3 & 1.3 & 1.20 & 1.26 & 1.3 & 1.4 & 1.4 & 1.4 & 1.3 & 1.2 \\
\hline
.8 & 1.4 & 1.2 & 1.3 & 1.22 & 1.31 & 1.4 & 1.5 & 1.5 & 1.5 & 1.4 & 1.4 \\
\hline
.9 & 1.1 & 1.5 & 1.7 & 1.88 & 1.70 & 1.7 & 1.7 & 1.7 & 1.6 & 1.6 & 1.6 \\
\hline
1.0 & 1 & 1 & 1 & 1 & 1 & 1 & 1 & 1 & 1 & 1 & 1 \\ \hline
\hline
\end{tabular}

%                 ----------- table 2b -----------

\newpage

\begin{tabular}{||c||c|c|c|c|c|c|c|c|c|c|c||} \hline \hline
$\sigma\backslash$ q  & $\frac{1}{16}$ & $\frac{1}{2}$ & 1 & 2 & 3 & 4 & 5 & 8
&
16 & 32 & 64 \\ \hline \hline
.1 & .11 & .12 & .12 & .101 & .091 & .12 & .16 & .21 & .31 & - & - \\ \hline
.2 & .18 & .20 & .21 & .202 & .215 & .24 & .30 & .45 & .75 & - & - \\ \hline
.3 & .32 & .31 & .31 & .301 & .323 & .41 & .53 & 1.1 & 4.1 & - & - \\ \hline
.4 & .34 & .34 & .35 & .373 & .481 & .59 & .89 & 3.0 & - & - & - \\ \hline
.5 & .33 & .34 & .37 & .430 & .577 & .77 & 1.2 & 5.0 & - & - & - \\ \hline
.6 & .30 & .32 & .37 & .501 & .664 & .83 & 1.2 & 2.3 & - & - & - \\ \hline
.7 & .25 & .29 & .35 & .518 & .636 & .78 & 1.0 & 1.4 & 4.6 & - & - \\ \hline
.8 & .18 & .24 & .32 & .483 & .574 & .67 & .80 & .99 & 1.3 & 1.4 & - \\ \hline
.9 & .10 & .21 & .30 & .405 & .491 & .56 & .62 & .76 & .97 & 1.1 & - \\ \hline
1.0 & .05 & .18 & .25 & .309 & .393 & .46 & .52 & .63 & .79 & .92 &
1.0 \\ \hline
\hline
\end{tabular}
%                ----------- table 3a -----------

\vskip .5in
\begin{tabular}{||c||c|c|c|c|c|c|c|c|c|c|c||} \hline \hline
$\sigma\backslash$ q  & $\frac{1}{16}$ & $\frac{1}{2}$ & 1 & 2 & 3 & 4 & 5 & 8
&
16 & 32 & 64 \\ \hline \hline
.1 & - & - & - & - & .94 & .89 & .84 & .83 & .61 & - & - \\ \hline
.2 & - & - & - & - & .75 & .84 & .76 & .56 & .24 & - & - \\ \hline
.3 & .72 & .74 & .75 & - & .78 & .65 & .50 & .24 & .04 & - & - \\ \hline
.4 & .71 & .72 & .72 & - & .52 & .48 & .30 & .08 & - & - & - \\ \hline
.5 & .72 & .72 & .71 & .70 & .48 & .37 & .23 & .05 & - & - & - \\ \hline
.6 & .75 & .74 & .72 & .49 & .40 & .36 & .24 & .12 & - & - & - \\ \hline
.7 & .78 & .77 & .77 & .43 & .46 & .41 & .30 & .24 & .06 & - & - \\ \hline
.8 & .81 & .92 & .99 & .43 & .59 & .57 & .47 & .42 & .31 & .32 & - \\ \hline
.9 & .87 & 1.00 & 1.00 & 1.41 & 2.16 & 1.32 & 1.08 & .84 & .62 & .62 & - \\
\hline
1.0 & 1 & 1 & 1 & 1 & 1 & 1& 1 & 1 & 1& 1 & 1 \\ \hline
\hline
\end{tabular}
%                  ----------- table 3b -----------

\vskip .5in

\begin{tabular}{||c||c|c|c||} \hline \hline
q & $\nu^{-1}_{e}$ & $\nu^{-1}_{TG}$ & $\Delta$ (\%) \\ \hline \hline
$\frac{1}{16}$ & .340 & .3250 & 4.5 \\ \hline
$\frac{1}{2}$ & .338  & .3300 & 2.5 \\ \hline
1 & .349 & .3387 & 3 \\ \hline
2 & .373 & .4 & 7 \\ \hline \hline
\end{tabular}
%                  ----------- table 4 -----------

%-------------------------------------------------------------------------
\end{document}